\documentclass[pra,twocolumn,eqsecnum,showpacs,amsmath,amssymb,superscriptaddress]{revtex4-1} 
\usepackage{graphicx,bm,color,mathptmx,hyperref,subfig}

\newcommand{\ZX}{\ensuremath{Z\!X}}

\begin{document}

\title{Practical implementation of mutually unbiased bases using quantum circuits} 

\author{U. Seyfarth} 
\affiliation{Max-Planck-Institut f\"ur die Physik des Lichts, 
G\"{u}nther-Scharowsky-Stra{\ss}e 1, Bau 24, 
91058 Erlangen, Germany}

\author{L. L. S\'{a}nchez-Soto} 
\affiliation{Max-Planck-Institut f\"ur die Physik des Lichts, 
G\"{u}nther-Scharowsky-Stra{\ss}e 1, Bau 24, 
91058 Erlangen, Germany}
\affiliation{Department f\"{u}r Physik, 
Universit\"{a}t Erlangen-N\"{u}rnberg, 
Staudtstra{\ss}e 7, Bau 2, 91058 Erlangen, Germany} 
\affiliation{Departamento de \'Optica, 
Facultad de F\'{\i}sica, Universidad Complutense, 
28040~Madrid, Spain}

\author{G. Leuchs} 
\affiliation{Max-Planck-Institut f\"ur die Physik des Lichts, 
G\"{u}nther-Scharowsky-Stra{\ss}e 1, Bau 24, 
91058 Erlangen, Germany}
\affiliation{Department f\"{u}r Physik, 
Universit\"{a}t Erlangen-N\"{u}rnberg, 
Staudtstra{\ss}e 7, Bau 2, 
91058 Erlangen, Germany}

\date{\today}

\begin{abstract}
  The number of measurements necessary to perform the quantum state
  reconstruction of a system of qubits grows exponentially with the
  number of constituents, creating a major obstacle for the design of
  scalable tomographic schemes.  We work out a simple and efficient
  method based on cyclic generation of mutually unbiased bases. The
  basic generator requires only Hadamard and controlled-phase gates,
  which are available in most practical realizations of these systems.
  We show how complete sets of mutually unbiased bases with different
  entanglement structures can be realized for three and four qubits.
  We also analyze the quantum circuits implementing the various
  entanglement classes.
\end{abstract}

\pacs{03.65.Wj, 03.65.Aa, 03.67.Ac, 03.67.Lx}

\maketitle

\section{Introduction}
\label{sec:intro}

Modern quantum science is nearing precise control and
manipulation of quantum states so as to achieve results beyond the
limits of conventional technologies.  Quantum-enhanced
devices are already on the market and point to a transformation
of measurement, communication, and computation.

For the successful completion of these tasks, verification of each
stage in the experimental procedures is of utmost importance; quantum
tomography is the appropriate tool for that
purpose~\cite{lnp:2004uq}. The main challenge of this
technique is simple to state: given a system in a state represented by
the density matrix $\varrho$ and an informationally complete
measurement~\cite{Prugovecki:1977fk,Busch:1989kx,Sych:2012il}, the
state $\varrho$ must be inferred from the distinct measurement
outcomes.
 
For a $d$-dimensional quantum system (a qudit, in the modern
parlance of quantum information) this amounts to determining
$d^{2} - 1$ independent real numbers. A von Neumann measurement (the
only ones we consider here) fixes at most $d - 1$ real parameters, so
$d + 1$ different tests have to be performed to reconstruct
the state.  This means that $d^{2} + d$ histograms have to be
recorded.  The approach is, thus, suboptimal because this number is higher
than the number of parameters in the density matrix.  This redundancy
is optimized when the bases in which the measurements are performed
are mutually unbiased~\cite{Ivanovic:1981ly,Wootters:1989kq}.

At a fundamental level, mutuallly unbiased bases (MUBs) are intimately
related to the nature of quantum information and provide the most
accurate statement of complementarity. The idea emerged in the
pioneering work of Schwinger~\cite{Schwinger:1960dq} and it has
gradually turned into a primitive of quantum theory: apart from the
role in quantum tomography, they are instrumental in addressing a
number of enthralling questions~\cite{Durt:2010cr}.

However, tomography becomes harder as we explore more
intricate systems. If we look at the simple, yet illustrative case of
$n$ qubits, even with MUBs, one will have to make at least
$2^n+1$ measurements before one can claim to know
everything about an \textit{a priori} unknown system.  With such a
scaling, it is clear that the methods rapidly become intractable for
present state-of-the-art experiments~\cite{Monz:2011lq,Yao:2012hc}. 

We are thus inevitably led to the quest for tomographical techniques
with better scaling. A promising class of new protocols are explicitly
optimized only for particular kinds of states.  This includes states
with  low rank~\cite{Gross:2010dq,Flammia:2012if,Guta:2012bl}, with 
special emphasis in some relevant cases as matrix
product (MPS)~\cite{Cramer:2010oq,Baumgratz:2013fq} ,or multiscale entangled
renormalization ansatz (MERA)
states~\cite{Landon-Cardinal:2012mb}. The specific but pertinent
example of permutationally invariant qubits has been also
examined~\cite{Ariano:2003qf,Toth:2010dq,Klimov:2013ya,Moroder:2012bs},
as they are of great import in diverse quantum information
strategies~\cite{Berry:2000qa,Stockton:2003fv,Bartlett:2003dz,
  Cabello:2007ij,Fiurasek:2009bs,Demkowiczi:2009kl,Hentschel:2011fu}.

In this paper, we devise an approach that puts a new spin on the
problem.  We revisit the MUB strategy, but capitalize on a recently
developed construction which generates the corresponding MUBs in a
cyclic way~\cite{Kern:2010lc,Seyfarth:2011ru}. From an experimental
viewpoint, the undeniable advantage of this approach is that a single
unitary operation $U$ is enough to create all the MUBs. Furthermore,
this single unitary operator can be expressed as a quantum circuits
involving exclusively Hadamard and controlled-phase
gates~\cite{Chuang:2000fk}. In this way, the number of gates scales
only linearly in the number of qubits, which is an optimal scaling.

Our paper is organized as follows: In Sec.~\ref{sec:prelim} we
concisely sketch the rudiments of our method. For systems of qubits, it
is well known that different complete sets of MUBs exist with distinct
entanglement properties~\cite{Lawrence:2002ij,Lawrence:2004xj,
  Romero:2005dz,Lawrence:2011fu,Wiesniak:2011kl,Rehacek:2013jt,Spengler:2012qa}.
In Sec.~\ref{sec:results} we work out the simple example of three
qubits, showing the quantum circuits associated to the different
complete sets, while the case of four qubits is worked out in the
Supplemental Material. Finally, our conclusions are briefly summarized in
Sec.~\ref{sec:conclusion}.

\section{Mutually unbiased bases: Basic background}
\label{sec:prelim}

We consider a $d$-dimensional quantum system with Hilbert space
isomorphic to $\mathbb{C}^d$. The different outcomes of a maximal test
constitute an orthogonal basis of
$\mathbb{C}^{d}$~\cite{Peres:1993bu}.  One can also look for
orthogonal bases that, in addition, are ``as different as possible''.
This is the idea behind MUBs and can be formally stated as follows:
two orthonormal bases
$\mathcal{B}_{j} = \{ | \psi_{\ell}^{(j)} \rangle \}$ and
$\mathcal{B}_{j^{\prime}} = \{ | \psi_{\ell^{\prime}}^{(j^{\prime})}
\rangle \}$ ($j \neq j^{\prime}$) are mutually unbiased when
\begin{equation}
  \label{eq:defMUBS}
  | \langle \psi_{\ell}^{(j)}  | 
  \psi_{\ell^{\prime}}^{(j^{\prime})} \rangle |^2 
 = \frac{1}{d} \, , 
  \qquad
  \forall \,  \ell,  \ell^{\prime}  = 1, \ldots, d \, .
\end{equation}
Unbiasedness also applies to measurements: two nondegenerate tests are
mutually unbiased if the bases formed by their eigenstates are MUBs.
For example, the measurements of the components of a spin 1/2 along
the $x$, $y$, and $z$ axes are all unbiased.

It has been shown that the number of MUBs is at most
\mbox{$d+1$}~\cite{Ivanovic:1981ly}, and that such a complete set
exists whenever $d$ is a prime or power of a
prime~\cite{Calderbank:1997ao}.  Remarkably, there is no known answer
for any other values of $d$, although there have been some attempts to
find a solution to this problem in some simple cases, such as
$d=6$~\cite{Grassl:2004iz,Butterley:2007fe,Brierley:2008ja,
  Brierley:2009mw,Raynal:2011xc,McNulty:2012ez} or when $d$ is a
non-prime integer squared~\cite{Archer:2005bz,Wocjian:2005pw}.

In what follows, we concentrate on a system of $n$ qubits, where the
dimension of the space is $d=2^{n}$. The basic single-particle Pauli operators
$\sigma_{z}$ and $\sigma_{x}$ are
\begin{equation}
  \sigma_{z} = | 1 \rangle \langle 1 | - |0 \rangle \langle 0 | \, ,
  \qquad \qquad
  \sigma_{x} = | 0 \rangle \langle 1 | + | 1 \rangle \langle 0 | \, ,
  \label{sigmas}
\end{equation}
where $|0 \rangle$ and $| 1 \rangle$ are the computational basis for a
single qubit. The concept can be extended to $n$ qubits by
introducing $2n$-dimensional vectors 
\begin{equation}
  \mathbf{a} = ( a^z_1,\ldots,a^z_n; a^x_1,\ldots, a^x_n)^{T} \, ,
\end{equation}
where $T$ denotes the transpose and
$a_{i}^{z}, a_{j}^{x} \in \mathbb{Z}_{2}$. In this way, the
generalized Pauli operators can be written down as
\begin{align}
\label{eqn:bandy:ZXa}
 \ZX (\mathbf{a}) = 
     (-i)^{a^{z}_{1} a^{x}_{1}}   \sigma_{z}^{a^z_1} \sigma_{x}^{a^x_1} 
   \otimes \dots \otimes
      (-i)^{a^{z}_{n} a^{x}_{n}} \sigma^{a^z_n} \sigma^{a^x_n} \, .
\end{align}
In technical jargon, this set is just the Weyl-Heisenberg group
(modulo its center). 

The importance of these operators lies in the observation
noticed in Ref.~\cite{Bandyopadhyay:2002db} that complete sets of MUBs
naturally arise  from a partition of the set of Pauli operators
into $d+1$ subsets of $d -1$  commuting operators, called
classes; they can be expressed as
\begin{align}
  \label{eqn:prelim:classes}
  \mathfrak{C}_j = \{ \ZX(\mathbf{a} ) : \mathbf{a} = G_j
  \mathbf{c} : \mathbf{c} \in \mathbb{Z}_2^n \} \, .
\end{align}
In this way,  each of the classes $\mathfrak{C}_j$ can be
 specified by the generator $G_j$.

 Within each class $\mathfrak{C}_j$ all Pauli operators commute.  If
 we unveil the tensor product of the Pauli operators, we can consider
 each Pauli operator as a joint operator that performs either a
 $\sigma_{z}$, $\sigma_{x}$, $\sigma_{y}$, or an identity operation on
 each single qubit separately.  Within a certain class, the Pauli
 operators on each qubit can either commute or not, which leads to
 different entanglement properties. The maximal entanglement occurs
 when the Pauli operators of one class commute only in combination,
 whereas no entanglement appears when they commute on every qubit
 separately. All possible partitions of the operators into their
 subsystems give rise to different entanglement properties, where a
 relabelling of the different sites should not influence this
 classification at all. Therefore, we define a vector $\mathbf{n}$
 which represents the entanglement structure of a certain set of MUBs:
 the entries of $\mathbf{n}$ are computed by counting the number of
 classes with each entanglement structure, starting from a completely
 factorizable system, and ending with a fully entangled one.

\begin{figure*}
\begin{center}
 \includegraphics[width=0.63\columnwidth]{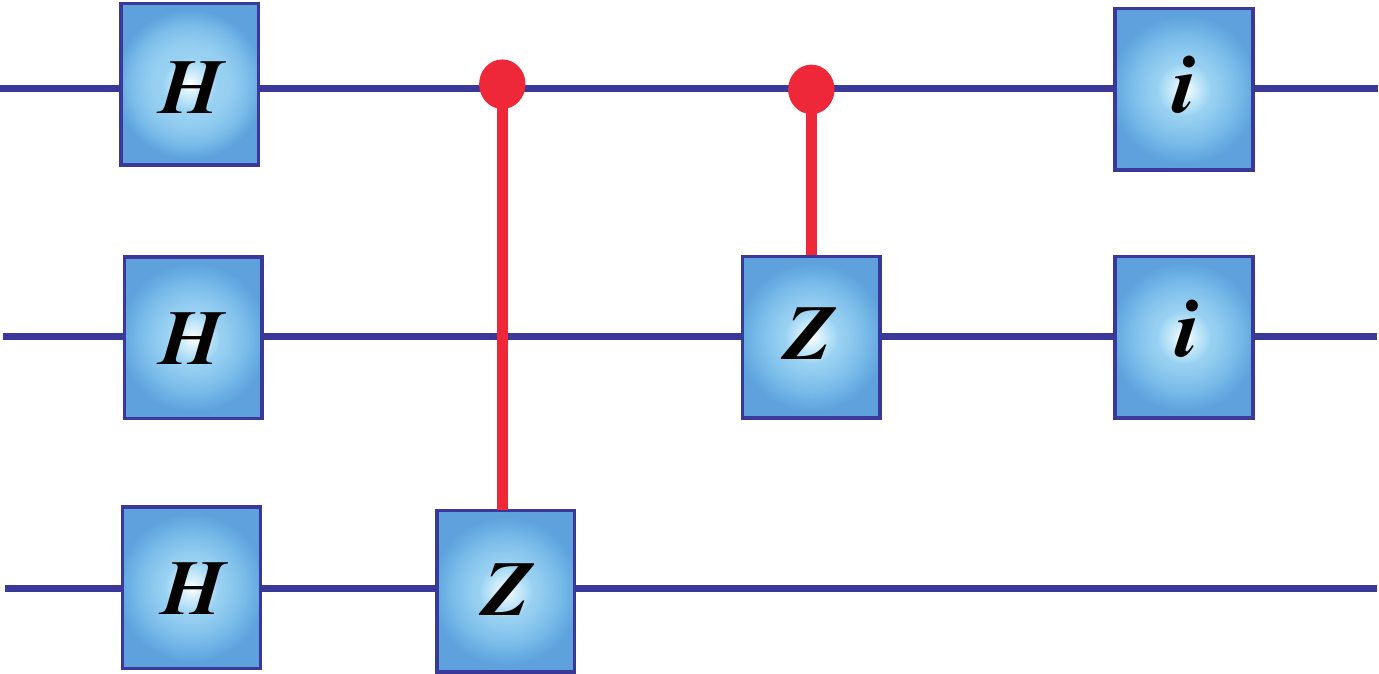} \qquad
 \includegraphics[width=0.63\columnwidth]{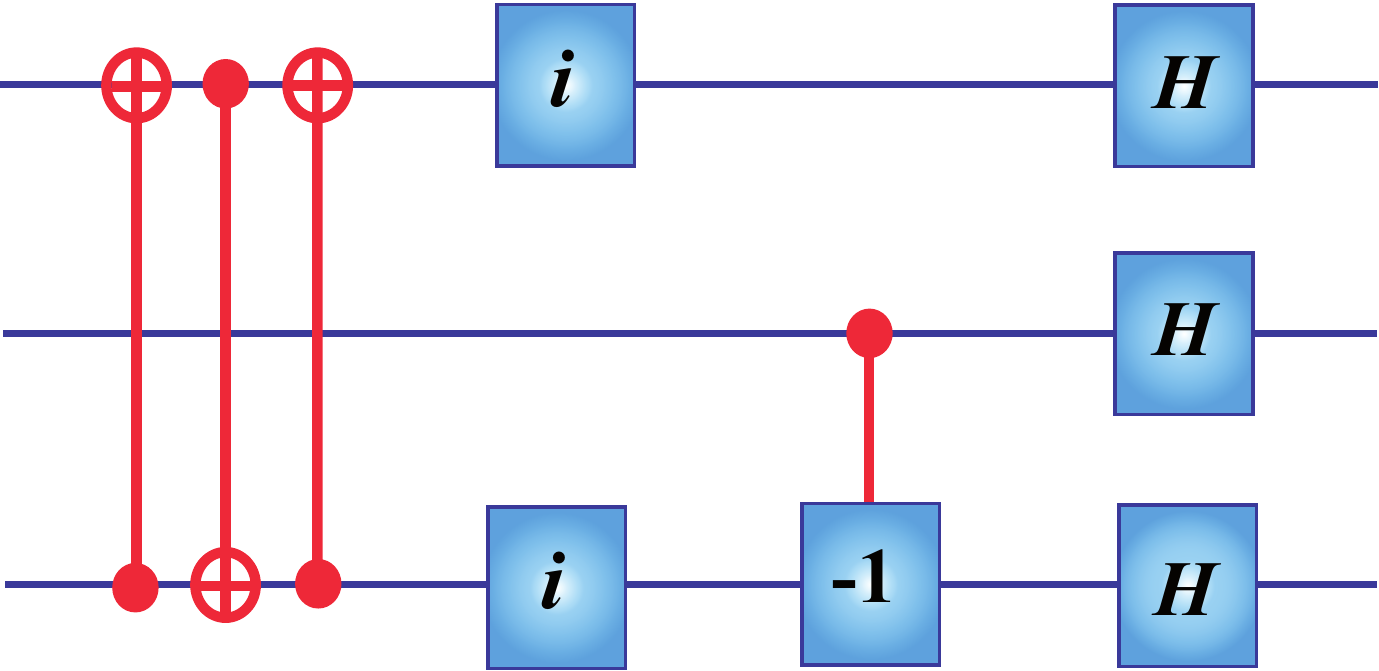} \qquad
 \includegraphics[width=0.63\columnwidth]{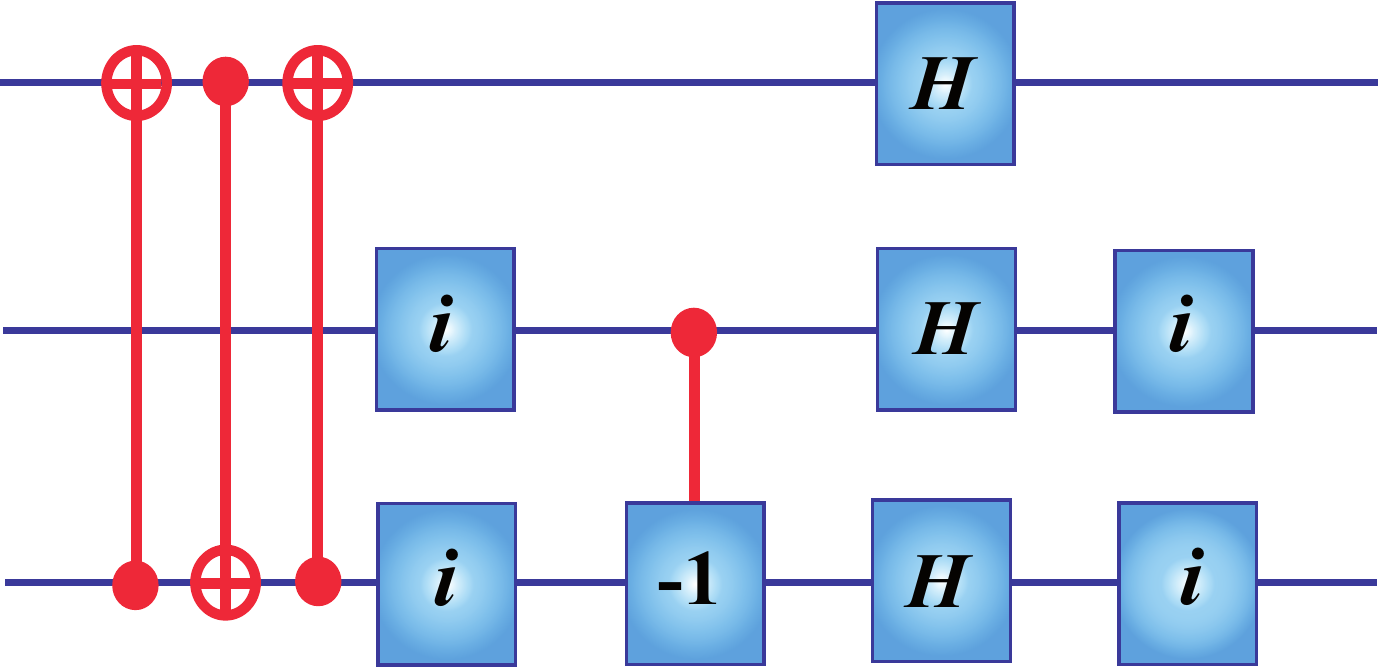} 
\end{center}
\caption{Quantum circuits implementing the generators of three-qubit
  MUBs with entanglement structures (from left to right) $(3,0,6)$,
  $(2,3,4)$, and $(1,0,6)$. The notation for the gates is the standard
  one~\cite{Chuang:2000fk}.}.
\end{figure*}

Different explicit constructions of MUBs in prime power dimensions
have been suggested in a number of recent
papers~\cite{Klappenecker:2003fk,Parthasarathy:2004kx,Pittenger:2004vn,
Durt:2005ys,Planat:2005zr,Klimov:2005oq,Boykin:2007ly}. We follow here
the approach established in Refs.~\cite{Kern:2010lc,Seyfarth:2011ru}, 
that allows a cyclic generation of the MUBs, that is, the generators
appearing in each class (\ref{eqn:prelim:classes}) can be 
expressed as
\begin{equation}
  \label{eq:gency}
  G_{j} = C^{j} G_{0} \, ,
\end{equation}
where $G_{0}$ is a fixed generator.  We skip  the mathematical
details involved in the  derivation of the method and content
ourselves with the final result, which looks very compact:
the symplectic matrix $C$ can be jotted down as
\begin{align}
 C = \begin{pmatrix} 
 B + A R^{-1} & R + BA +AR^{-1} A\\ 
R^{-1} & R^{-1} A
\end{pmatrix} \, ,
\end{align}
where $B$, $R$, and $A$ are $n \times n$ matrices whose properties
will be specified soon.  The successive powers of  $C$ can be computed
as 
\begin{widetext}
\begin{equation}
 C^j = 
\begin{pmatrix} 
F_{j+1}(B) + A R^{-1} F_{j} (B) & 
F_{j+1}(B) A + F_{j} (B) R +AR^{-1}[F_{j} (B) A +F_{j-1}(B) R] \\ 
R^{-1} F_{j} (B) & R^{-1}[ F_{j}(B) A+F_{j-1}( B) R]  
\end{pmatrix} \, .
\end{equation} 
\end{widetext}
Here $F_{j} (x) $ refer to the  Fibonacci polynomials, which  are a
generalization of the well-known Fibonacci sequence. They are defined
recursively as   
\begin{align}
 F_{j+1} (x) = x F_{j} (x) + F_{j-1} (x) \, ,
\end{align}
with  $F_0(x)=0$ and $F_1(x)=1$ and the coefficients therein
are binary numbers in $\mathbb{Z}_{2}$. In many considerations in this
work, we will take as the seed generator $G_0 = (\openone_{n},
0_{n})^t$,  which leads to 
\begin{align}
 G_{j} =
\begin{pmatrix} 
F_{j+1}(B) F^{-1}_j(B)  R + A \\ 
\openone_m\end{pmatrix} \, ,
\qquad
1 \leq j \leq d \, .
\end{align}

To ensure that  complete set of MUBs are generated, we have to impose
additional conditions. The first one, of rather technical
character, implies that the Fibonacci index~\footnote{The Fibonacci
  index of an irreducible polynomial $p (x)$ is the minimum integer
  $n$ such that $p(x)$ divides $F_n(x)$}.
of the characteristic
polynomial of $B$ has to be $d+1$. In addition,  $R$, $BR$, and $A$ have
to be symmetric and $R$ has to be invertible~\cite{Seyfarth:2014qd}.

It turns out  that when $R=\openone_m$ and $A=0_m$, the resulting
complete sets exhibit an entanglement structure with three completely
factorizable classes, which, following the original
work~\cite{Seyfarth:2014qd},  will be called field-based sets, as the
generators represent a finite field.  When $R$ is not a polynomial in
$B$ and $A=0_m$, the generators form an additive group, 
where  for only two of their classes the Pauli operators commute on
each qubit separately:  they are denoted as group-based sets, Finally,
whenever $R$ is not a polynomial in $B$, and $A$ is not the product of any
polynomial in $B$ with $R$ added to a diagonal matrix, the resulting
cyclic set of MUBs has only a single class left, where the Pauli
operators commute on all qubits separately. This case is denoted as
semigroup-based sets, as the generator represents an additive
semigroup.

\section{Results}
\label{sec:results}

The three-qubit system is the first nontrivial instance one can
consider, and  any complete set of MUBs exhibits \mbox{$2^3+1=9$} different
bases. It is well known~\cite{Lawrence:2002ij,Romero:2005dz,
Lawrence:2011fu,Garcia:2010rm} that each complete set of MUBs
possesses one  of the four different
entanglement structures, either $(3,0,6), (2,3,4), (1,0,6)$, 
or $(0,9,0)$. In this particular example, in $\mathbf{n} = (n_1,n_2,n_3)$,
$n_{1}$ denotes the number of separable bases (every eigenvector
of theses bases is a tensor product of singe-qubit states), $n_{2}$
the number of biseparable bases (one qubit is factorized and the other
two are in a maximally entangled state) and $n_{3}$ the number of
nonseparable bases.

To work out the cyclic construction of these sets, we first notice
that the only polynomial of order 3 that has full
Fibonacci index (i.e., index 9) is
\begin{align}
  p(x)=1+x+x^3.
\end{align}
 
For field-based sets, the matrix $B$ has to be
symmetric, as $R= \openone_m$.  The only possible
solution is
\begin{align}
  B=\begin{pmatrix}
    1 & 1 & 1 \\
    1 & 1 & 0 \\
    1 & 0 & 0
  \end{pmatrix} \, ,
\end{align}
or one of its permutations. This corresponds to an entanglement
structure $\mathbf{n} = (3,0,6)$.

The group-based sets are richer, as polynomials of $B$ can be shifted into $R$. One
possible solution is generated by
\begin{align}
 B = & 
\begin{pmatrix}
0&1&1\\
0&0&1\\
1&0&0
\end{pmatrix}  \, , 
\qquad
R = 
\begin{pmatrix}
0&0&1\\
0&1&0\\
1&0&0\end{pmatrix},
\end{align}
which leads finally to the symplectic matrix
\begin{align}
 C=\begin{pmatrix}
0&1&1&0&0&1\\
0&0&1&0&1&0\\
1&0&0&1&0&0\\
0&0&1&0&0&0\\
0&1&0&0&0&0\\
1&0&0&0&0&0
\end{pmatrix} \, .
\end{align}
This corresponds to the entanglement structure $\mathbf{n} =(2,3,4)$.

In a similar way, we find the following solution for the
semigroup-based sets
\begin{align}
 B=&
\begin{pmatrix}
1&1&1\\
1&1&0\\
1&0&0
\end{pmatrix} \, , 
\quad
R=\begin{pmatrix}
1&1&1\\
1&1&0\\
1&0&0
\end{pmatrix} \, , 
\quad
 A=
\begin{pmatrix}
1&0&1\\
1&1&0\\
1&0&0\end{pmatrix} \, ,
\end{align}
which gives the  matrix
\begin{align}
 C=\begin{pmatrix}
0&0&0&1&1&1\\
1&0&0&0&1&1\\
1&0&1&1&0&1\\
0&0&1&1&0&0\\
0&1&1&0&1&0\\
1&1&0&0&1&1
\end{pmatrix}.
\end{align}
and the corresponding entanglement structure is $\mathbf{n}=(1,6,2)$.

The set  $\mathbf{n}=(0,9,0)$ cannot be worked out initially 
from this construction method. However, this can be easily fixed:  as this set
does not contain any basis that measures properties of a completely 
factorizable system, a sort of offset operation transforming the
standard basis is needed. Therefore, the generator $G_0$ cannot be
taken as $(\openone_m,0_m)$ anymore, but instead its $X$-part, which
is $0_m$,  has to  be replaced with 
\begin{align}
\label{eq:pof}
 G_0^x=
 \begin{pmatrix}
   0&1&0\\
   1&0&0\\
   0&0&0
 \end{pmatrix},
\end{align}
and so  
\begin{align}
C = 
\begin{pmatrix}
0 & 0 & 0 & 0 & 0 & 1\\
1 & 0 & 0 & 1 & 0 & 0\\
0 & 0 & 0 & 0 & 1 & 0\\
0 & 0 & 1 & 0 & 0 & 0\\
1 & 0 & 0 & 0 & 0 & 0\\
0 & 1 & 0 & 0 & 0 & 0
\end{pmatrix} \, .
\end{align}

\begin{figure}
\begin{center}
 \includegraphics[width=\columnwidth]{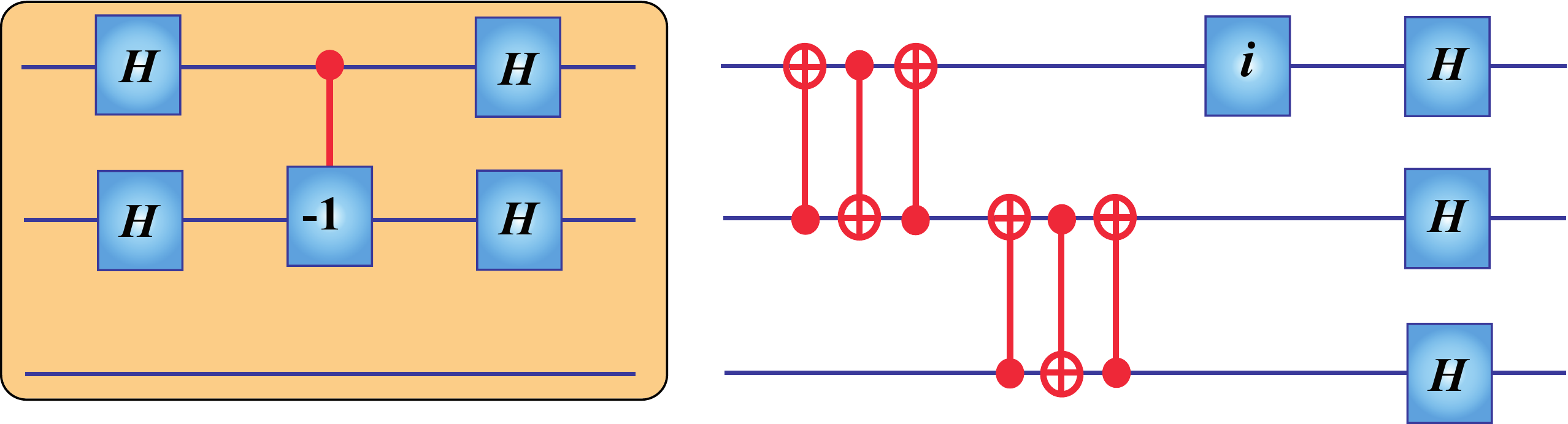} 
\end{center}
\caption{Quantum circuit implementing the generator of three-qubit
  MUBs with entanglement structure $(0,9,0)$. In the left, enclosed in
  a box, we show the circuit for the offset generator $G_0^x$.}
\end{figure}
and for the implementation of the symplectic generator

One of the outstanding advantages of our approach is that the unitary generator can
be worked out in quite a direct way as a quantum circuit involving only
elementary gates. Such a decomposition can be immediately found
following the standard rules~\cite{Chuang:2000fk}. In particular, this
is relevant for a practical implementation.  In Fig.~1 we
summarize the circuits corresponding to the structures
$(3,0,6), (2,3,4), (1,0,6)$, whereas in Fig.~2 we give the circuit
for $(0,9,0)$, including the offset (\ref{eq:pof}).

The method works for any number of qubits. Since the
ideas are analogous, we omit the unnecessary details, although, for
completeness, we give the complete solution for four qubits in the
Supplemental Material.

\section{Conclusions}
\label{sec:conclusion}

In short, we have shown the construction of cyclic MUBs for $n$ qubits
with all possible entanglement structures. On physical grounds, one
could expect that the performances of these different classes with
respect to entanglement-specific state properties will also be
different. In our approach, this is reflected in the different
complexities of the associated generator.  Finally, the fact that only
one generator needs to be implemented to generate the whole set of
MUBs makes this method especially interesting and a potential
candidate for a realistic scheme for current experimental setups.

\begin{acknowledgments}
  We thank Olivia di Matteo for fruitful discussions.  Financial
  support from the EU FP7 (Grant Q-ESSENCE), the Spanish DGI (Grant
  FIS2011-26786) and Program UCM-Banco Santander (Grant GR3/14) is
  gratefully acknowledged.
\end{acknowledgments}


%

\end{document}